# AN EXPERIMENT
# WITH A USER MANUAL
# BASED ON
# A DENOTATIONAL SEMANTICS

(working version)

Andrzej Blikle

May 28th, 2019

## Abstract

Denotational models should provide an opportunity for the revision of current practices seen in the manuals of programming languages. New styles should on one hand base on denotational models but on the other — do not assume that today readers are acquainted in this field. A manual should, therefore, provide some basic knowledge and notation needed to understand the definition of a programming language written in a new style. At the same time — I strongly believe that — it should be written for professional programmers rather than for amateurs. The role of a manual is not to teach the skills of programming. Such textbooks are, of course, necessary, but they should tell the readers what the programming is about rather than the technicalities of a concrete language. The paper contains an example of a manual for a virtual programming language Lingua developed in our project.

**An invitation to project** This paper has been prepared within a project *Denotational Engineering* described on:
http://www.moznainaczej.com.pl/denotational-engineering/denotational-engineering-eng.
Persons interested to join the project are invited to see:
http://www.moznainaczej.com.pl/an-invitation-to-the-project

# Contents

# 1 Introduction

Denotational models should provide an opportunity for the revision of current practices seen in the manuals of programming languages. New styles should on one hand base on denotational models but on the other — do not assume that today readers are acquainted in this field. A manual should, therefore, provide some basic knowledge and notation needed to understand the definition of a programming language written in a new style. At the same time — I strongly believe that — it should be written for professional programmers rather than for amateurs. The role of a manual is not to teach the skills of programming. Such textbooks are, of course, necessary, but they should tell the readers what the programming is about rather than the technicalities of a concrete language.

A virtual programming language **Lingua** was defined in [14] as an exercise of developing (designing) a language in "reverse order," i.e. from denotations to syntax[1]. This method permits the development of a language with two following attributes:

1. a full and formal denotational semantics,

2. a set of program constructors that guarantee the development of totally-correct programs with clean termination[2].

Ones we have a language with denotational semantics, we can define program-construction rules that guarantee the correctness of developed programs. This method consists in developing so-called *metaprograms*, i.e. programs that syntactically include their specifications. The method guarantees that if we compose two or more correct programs into a new program or if we transform a correct program, we get a correct program again. The correctness proof of a program is hence implicit in the way the program has been developed[3].

Besides its contribution to the development of correct programs, denotational semantics has one more important advantage: it provides a mathematical fundament for a comprehensive, complete and compact manual of the language. The present paper is an attempt to justify this claim by showing a sketch of a manual for **Lingua**.

In constructing **Lingua,** I assumed three basic priorities that determined the choice of programming mechanisms:

- the priority of the mathematical simplicity of the model; e.g., this resulted in the resignation from `goto` instructions,

- the priority of the simplicity of program construction rules; e.g., the assumption that the declarations of variables and procedures, as well as the definitions of types, should always be located at the beginning of a program,

- the priority of protection against "oversight errors" of a programmer; e.g., the resignation of any side-effects in procedures.

---

[1] That paradigm was introduced and studied in [8], [10], [11] and [13].
[2] Clean termination means that the program neither loops nor aborts with an error massage. This issue was investigated in [6].
[3] That paradigm was introduced and studied in [2], [4], [5] and [6].

All these commitments forced me to give up certain programming constructions which — although denotationally definable — would lead to complicated descriptions and even more complicated program-construction rules. In the sequel, such decisions will be referred to as *engineering decisions*.

Complementary to engineering decisions are *mathematical decisions* forced by the denotationality of our model. For instance, there are no procedures in **Lingua** that may take themselves as actual parameters[4]. There are also no concurrency mechanisms since the development of a "fully" denotational semantics for such a language (if at all possible) would require separate research[5].

At the level of data structures, **Lingua** covers Booleans, numbers, words, records, arrays and their arbitrary combinations plus SQL databases. It is also equipped with a relatively rich mechanism of types, e.g. covering SQL-like integrity constraints, and with tools allowing the user to define his/her own types structurally. At the imperative level, **Lingua** contains structured instructions, type definitions, procedures with recursion and multi-recursion.

Of course, **Lingua** is not a real language since otherwise [14] would become unreadable. It only illustrates the method which (hopefully) may be used in designing a real language in some future. For the same reason, the present paper is not a full manual of our language. It is restricted to giving guidelines of how to write such a manual under the assumption that it is addressed to professional programmers with some mathematical background, rather than to amateurs who have to be explained what programming is about[6].

My sketch of a manual consist of three parts:

1. An introduction to a specific mathematical notation and basic mathematical concepts of denotational models (Sec.2).

2. A half-formal introduction to the syntax and the semantics of the language illustrated by examples (Sec. 3 and Sec.4).

3. A full formal definition of syntax and semantics (Sec.5).

Although I assume that the manual is addressed to professional programmers, I do not expect them to be familiar with the ideas of denotational semantics. These ideas and the corresponding notation are therefore introduced in the first part.

The second part is devoted to an intuitive — although mathematical — explanation of basic concepts and constructions of the language. After having read this part, the reader should understand the general structure and the philosophy of the language and should be able to read the third part with sufficient understanding.

The third part is a glossary of syntactic and semantic constructors with their definitions. It may be skipped in the first reading remaining a reference source when it comes to writing programs.

---

[4] Such procedures were allowed in Algol 60, a popular programming languages in the years 1960-1980. A denotational model for such procedures (constructed by Dana Scott) is not expressible is "usual" set-theory where a function cannon be applied to itself. For more see Sec.4.1 in [14].

[5] There exist mathematical semantics of concurrency which can be said to be only "partially denotational". An example of such a solution is a "component-based semantics" (cf. [1]), where the denotations of programs' components are assigned to programs in a compositional way (i.e. the denotation of a whole is a composition of the denotations of its parts), but the denotations themselves are so called fucons whose semantics is defined operationally.

[6] As a matter of fact I could never understand why the authors of manuals of even such advanced languages like e.g. Phyton or Delphi assume that their readers know nothing about programming.

Due to this philosophy, a manual can be made relatively readable without affecting its mathematical precision and completeness.

However, to keep my paper of a size acceptable for publication, I skip formal definitions of semantics in the third part. I also omit SQL mechanisms which are technically rather complicated, and on the other hand, their discussion would not contribute much to the issue of how to write manuals. Readers interested in these subjects may find them in [14].

To make the present paper possibly self-contained I summarize in Sec.2 below all mathematical preliminaries which are necessary to read the paper and which have been described in detail in [14]. Besides, since our discussion about manuals is illustrated on the example of a virtual language defined in [14], all formal definitions of syntax have been 1-1 copied from there.

# 2  Mathematical preliminaries

For a full description of mathematical tools used in the development of denotational models see Sec.2 of [14]. Below is a selection of concepts that are used in the present paper. The notation bases on a metalanguage **MetaSoft** that was developed in the decade of 1980. in the Institute of Computer Science of the Polish Academy of Sciences (see [7]).

## 2.1    Notational conventions

The goal of my experiment is to write a manual based on denotational semantics but without too much of abstract mathematics. If in some future denotational models gain (hopefully) the acceptance of the community of programmers, the manuals will be written with their full mathematical content. In this manual, however, I do not assume that the reader is acquainted with [14] and therefore I use only as much of a mathematical language as is necessary to make the paper sufficiently clear and concise. Let me start with some basic notations:

- a : A means that a is an element of the set A; according to the denotational dialect "sets" are most frequently called "domains",

- f.a denotes f(a), and f.a.b.c denotes ((f(a))(b))(c); intuitively f takes a as an argument and returns the value f(a) which is a function which takes b as an argument and returns the value (f(a))(b), which is again a function…

- A → B denotes the set of all *partial functions* from A to B, i.e., functions possibly undefined for some elements of A,

- A ↦ B denotes the set of all *total functions* from A to B, i.e., functions undefined for all elements of A; of course, each total function is a particular case of a partial function, i.e. A ↦ B is a subset of A → B,

- A ⇒ B denotes the set of all finite function from A to B, i.e. functions defined for only finite subsets of A; such functions are called *mappings,* and of course, each mapping is a particular case of a partial function,

- A | B denotes the set-theoretic union of A and B,

- A x B denotes the Cartesian product of A and B,

- $A^{c*}$ denotes the set of all finite (possibly empty) tuples of the elements of A,

- $A^{c+}$ denotes the set of all finite non-empty tuples of the elements of A,

- tt and ff denote logical values „true" and „false" respectively,

- many-character symbols like dom, bod, com denote metavariables running over domains and if they are written with quotation marks as 'abdsr' denote themselves, i.e., metaconstants[7].

- in the definitional clauses of **Lingua** instead of indexed variables like $sta_1$, we write sta1 or sta-1 which is closer to a notation used in programs.

In this paper three different linguistic levels are distinguished:

1. the level of the basic text written in plain English and typed in Times New Roman,

2. the level of a formal, but not formalized, metalanguage **MetaSoft** whose formulas will be written with Arial,

3. the level of formalized programming language **Lingua** whose syntax, i.e. programs and their components, will be written in Courier New.

The difference between "formal" and "formalized" is such that the former is introduced intuitively as mathematical notation, whereas the latter requires an explicit definition of syntax (usually by a grammar) and a formal definition of semantics.

A frequently used construction in **MetaSoft** is a *conditional definition of a function* with the following scheme:

f.x =

    $p_1$.x ➜ $g_1$.x

    $p_2$.x ➜ $g_2$.x

    …

    **true** ➜ $g_n$.x

where each $p_i$ is a classical predicate, i.e., a total function with logical values tt or ff, **true** is a predicate which is always satisfied, and each $g_i$ is just a function. The formula above is read as follows:

if $p_1$.x is true, then f.x = $g_1$.x and otherwise,

if $p_2$.x is true, then f.x = $g_2$.x and otherwise,

…

and otherwise $g_n$.x.

Intuitively speaking the evaluation of such a function goes line by line and stops at the first line where $p_i$.x is satisfied.

In the scheme above I also allow the situation where in the place of a $g_i$.x we have the undefinedness sign "?" which means that for x that satisfies $p_i$.x the function f is undefined. This convention is used in conditional definitions of partial functions.

In conditional definitions we also use a technique similar to defining local constants in programs. For instance if f : A x B ⟼ C we can write

f.x =

---

[7] Metavariables and metaconstants are objects of the metalanguage **MetaSoft** whereas variables and constants are objects of the programming language **Lingua**.

$p_1.x \qquad \Rightarrow g_1.x$

**let**

$\quad$ (a, b) = x

$p_2.a \Rightarrow g_2.x$

$p_3.b \Rightarrow g_3.x.$

which is read as: "let x be a pair of the form (a, b)". We can also use **let** in the following way:

f.x =

$\quad p_1.x \qquad \Rightarrow g_1.x$

$\quad$ **let**

$\qquad$ y = h.x

$\quad p_2.x \Rightarrow g_2.y$

$\quad p_3.x \Rightarrow g_3.y.$

All these explanations are certainly not very formal, but the notation should be clear when it comes to concrete examples in the sequel of the paper.

By $[a_1/v_n,\ldots,a_n/v_n]$ we denote a finite-domain functions with domain $\{a_1,\ldots,a_n\}$ and the corresponding values $\{v_1,\ldots,v_n\}$. By $f[a_1/v_n,\ldots,a_n/v_n]$ we denote an overwriting of f by $[a_1/v_n,\ldots,a_n/v_n]$, i.e. a function which differs from f only on the domain $\{a_1,\ldots,a_n\}$.

For any two functions $f : A \rightarrow B$ and $g : B \rightarrow C$ by $f \bullet g$ we mean the *sequential composition* of these functions, i.e.

$(f \bullet g).a = g.(f.a)$

## 2.2    Many-sorted algebras

The denotational model of a programming language investigated in [14] is based on the concept of a *many-sorted algebra*. Half formally, a many-sorted algebra is a finite collection of sets, called the *carriers* of the algebra, and a finite collection of functions called the *constructors* of the algebra. The constructors take arguments from carriers and return their values to carriers. A graphical representation of a two-sorted algebra of numbers and Booleans is shown in Fig. 2.2-1. This algebra will be referred to as NumBool.

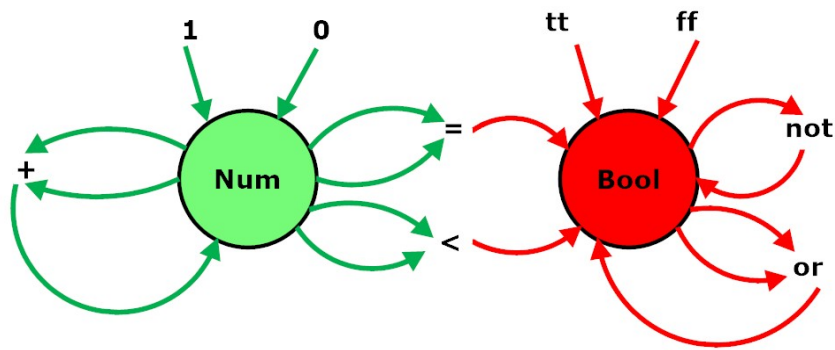

**Fig. 2.2-1 Graphical representation of a two-sorted algebra** NumBool

A textual representation of NumBool — called the *signature* of this algebra — is shown on the left part of Fig. 2.2-2.

In our algebra, we have four zero-argument constructors 1, 0, tt, ff, one one-argument constructor not, and four two-argument constructors +, =, <, or. The zero-argument constructors create elements of carriers "from nothing", whereas all other constructors create elements of carriers from other elements of carriers.

An element of an algebra is called *reachable* if it can be constructed (reached) using the constructors of the algebra. In NumBool, where Num denotes the set of <u>all</u> real numbers, the *reachable subset* of Num contains only non-negative integers.

By a *reachable subalgebra* of an algebra we mean its subalgebra with carriers restricted to their reachable parts. In our case, this is an algebra of nonnegative integers and Booleans.

| The algebra NumBool | | | The algebra NumBoolExp | | |
|---|---|---|---|---|---|
| 1 | : | ↦ Num | 1 | : | ↦ NumExp |
| 0 | : | ↦ Num | 0 | : | ↦ NumExp |
| + | : Num x Num | ↦ Num | + | : NumExp x NumExp | ↦ NumExp |
| = | : Num x Num | ↦ Bool | = | : NumExp x NumExp | ↦ BoolExp |
| < | : Num x Num | ↦ Bool | < | : NumExp x NumExp | ↦ BoolExp |
| tt | : | ↦ Bool | tt | : | ↦ BoolExp |
| ff | : | ↦ Bool | ff | : | ↦ BoolExp |
| not | : Bool | ↦ Bool | not | : BoolExp | ↦ BoolExp |
| or | : Bool x Bool | ↦ Bool | or | : BoolExp x BoolExp | ↦ BoolExp |

**Fig. 2.2-2 The signatures of two mutually similar algebras**

An algebra is said to be *reachable* if all of its carriers contain only reachable elements. Notice that if we remove the zero-argument constructor 1 from our algebra, then the reachable subset of Num would be empty.

In the algebraic approach to denotational models the algebra of programs meanings, i.e. the *denotations*, are usually unreachable, whereas the algebras of syntax are reachable by definition (this should become clear in Sec.2.3).

On the right-hand side of Fig. 2.2-2 we have the signature of a syntactic algebra Num-BoolExp of (variable-free) expressions. This algebra is *similar* to NumBool in the sense that there is a one-one correspondence between the constructors and carriers of both algebras (for a formal definition see Sec.2.11 of [14]). In our case this correspondence is implicit in the notation: 1 corresponds to 1, 0 corresponds to 0, NumExp corresponds to Num etc. The constructors of NumBoolExp create expressions. E.g. the constructor + given two expressions exp-1 and exp-2 creates the expression[8]:

    +(exp-1, exp-2)

Examples of expressions are:

    1, 0, +(1,1), +(1,+(1,0)), not(<(1,+(1,1))

Now assume that NumExp and BoolExp contain only reachable expressions, i.e. that Num-BoolExp is a reachable algebra. In that case, we may define two functions

SemE : NumExp ↦ Num

---

[8] For simplicity I use here the same symbol "+" to denote a constructor of expressions and a syntactic symbol of addition.

SemB : BoolExp ⟼ Bool

called respectively the *semantics of numerical expressions* and the *semantics of Boolean expressions* whose definitions are inductive relative to the structure of expressions:

SemE.[`1`] = 1

SemE.[`+(exp-1, exp-2)`] = SemE.[`exp-1`] + SemE.[`exp-2`]

etc.

For instance :

SemE.[`+(1,+(1,0))`] = 2

SemB.[`<(+(1,+(1,0)),0)`] = ff

Notice that both functions are "gluing" many different expressions into the same numbers resp. Boolean element, e.g.

SemE.[`+(1,+(1,0))`] = SemE.[`+(1,1)`] = 2

SemB.[`<(+(1,+(1,0)),0)`] = SemB.[`<(0,0)`] = ff

In the algebraic language, the tuple of our semantics (SemE, SemB) is called a *homomorphism* from NumBool into NumBoolExp. In the language of denotational models, this homomorphism is called the *semantics* of the two-sorted language of expressions identified by the algebra NumBoolExp. In this case, numbers are the *denotations* of numeric expressions, and tt and ff are the *denotations* of Boolean expressions.

It is clear from our example that a homomorphism may exist only between two similar algebras. Of course, this is only a necessary condition (more of that issue in see Sec.2.13 in [14]).

## 2.3    Equational grammars

Let A be an arbitrary finite set of symbols called an *alphabet*. By a *word* over A, we mean every finite sequence of the elements of A including the empty sequence. Traditionally words are written as sequences without commas between the characters, e.g., accbda and the *empty word* is denoted by ε.

If x and y are words, then by their *concatenation* — which we denote by x © y or simply by xy — we mean a sequential combination of these words. E.g.

abdaa © eaag = abdaaeaag

Also, the function © itself is called *concatenation*. Every set L of words over A is called a *formal language* (or simply a *language*) over A. By Lan(A) we denote the family of all languages over A and by Ø — the empty language (empty set). If P and Q are languages, then their *concatenation* is the language defined by the equation:

P © Q = {p © q | p:P **and** q:Q}.

As we see, by © we denote not only a function on words but also on languages. If it does not lead to ambiguities, P © Q is written as PQ. Since concatenation is an associative operation, we can write PQL instead of (PQ)L or P(QL). We shall also assume that concatenation binds stronger than set-theoretic union, hence instead of

(P © Q) | (R © S)

we shall write

PQ | RS

It is also easy to see that concatenation is distributive over the union, i.e.

(P | Q) R = PR | QR.

The n-th *power of a language* P is defined recursively:

$P^0 = \{ \varepsilon \}$

$P^n = P \, \copyright \, P^{n-1}$ for $n > 0$

Another two operators on languages are called respectively *plus* and *star*:

$P^+ = U.\{P^i \mid i > 0\}$

$P^* = P^+ \mid P^0$

Hence for an alphabet A, the set $A^+$ is the set of all non-empty words over A, and $A^*$ is the set of all words over A. Languages over A are subsets of $A^*$.

By an *equational grammar* over an alphabet A we mean a set of recursive equations of the form:

$X_1 = p_1.(X_1,\ldots,X_n)$

…

$X_n = p_n.(X_1,\ldots,X_n)$

where $X_i$'s run over languages over A and all $p_i$'s are operations on languages constructed as combinations of union, concatenation, power, star and plus operations[9]. As a tool for defining languages they are equivalent to the well-known context-free grammars, i.e. they define all and only context-free languages.

Every equational grammar defines unambiguously a reachable algebra of words. Consider the following grammar of numeric and Boolean expressions without variables:

NumExp = `0` | `1` | +(NumExp, NumExp)

BoolExp = `tt` | `ff` | =(NumExp, NumExp) | <(NumExp, NumExp) |
          `not`(BoolExp) | `or`(BoolExp, BoolExp)

According to a usual style for writing grammars `0`, `1`, `tt`, `ff`, `+`, `=`, `<`, `not`, `or`, `(`, `)` and the coma denote one-element languages: {0}, {1},…

By definition, the algebra corresponding to this grammar is the reachable subalgebra of NumBoolExp defined in Sec.2.2. Its carriers coincide with the languages defined by our grammar.

## 2.4     A denotational model of a programming language

In our approach, a denotational model of a programming language is a diagram of four algebras as shown in Fig. 2.4-1.

---

[9] Equational grammars were formally introduced and investigated in [2].

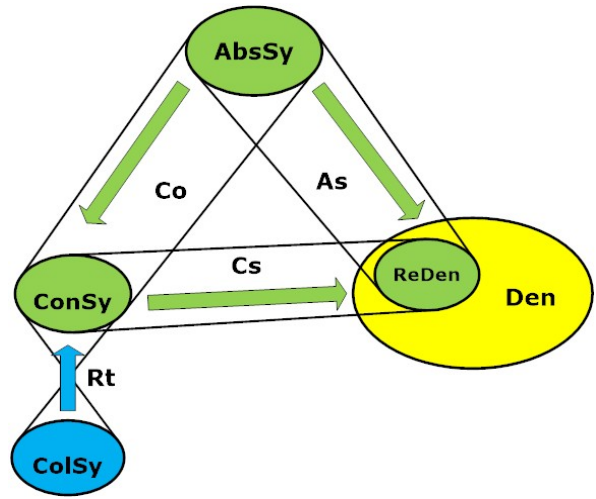

**Fig. 2.4-1 A denotational model of a programming language**

In this model:

Den        is an algebra of denotations of the language; usually its elements are functions that map states to states or states to values,

ReDen      is the reachable subalgebra of Den; here we have the denotations of expressions, instructions, declarations etc.

AbsSy      this algebra is historically named an *abstract syntax* and intuitively is an algebra of parsing trees of programs' components; this algebra is constructed in such a way that the homomorphism As : AbsSy ↦ ReDen exists and is unique,

ConSy      this algebra is historically named a *concrete syntax* and is the syntactic algebra of our programming language of expressions, instructions, declarations etc.; this algebra is constructed in such a way that there exist two homomorphisms

Co : AbsSy ↦ ConSy    which maps parsing trees into programs' components

Cs : ConSy ↦ ReDen    which maps programs' components into their denotations

ColSy      this algebra is called *colloquial syntax* since it introduces some shortcuts, called *colloquialisms* into the concrete syntax; usually, there is no homomorphism between colloquial syntax and concrete syntax, but there is a programmable function

Rt : ColSy ↦ ConSy

which we call a *restoring function* which transforms colloquialisms into concrete forms.

The syntax available to programmers is the "union" of concrete and colloquial syntax which means that programmers may use both. The concrete syntax is defined by an equational grammar. The full grammar for **Lingua** is shown in Sec.5.1.

## 2.5   Abstract errors

For practically all expressions appearing in programs their values in some circumstances cannot be computed "successfully". Here are a few examples:

- the value of x/y cannot be computed if y = 0,

- the value of the expression x+1 cannot be computed if x has not been declared in the program,

- the value of x+y cannot be computed if the sum exceeds the maximal number allowed in the language,

- the value of the array expression a[k] cannot be computed if k is out of the domain of array a,

- the query "Has John Smith retired?" cannot be answered if John Smith is not listed in the database.

In all these cases a well-designed implementation should stop the execution of a program and generate an error message.

To describe that mechanism formally, we introduce the concept of an *abstract error*. In a general case abstract errors may be anything, but in our models, they are going to be texts, e.g. 'division-by-zero'. They are closed in apostrophes to distinguish them from metavariables at the level of MetaSoft.

The fact that an attempt to evaluating x/0 raises an error message can be now expressed by the equation:

x/0 = 'division-by-zero'

In the general case with every domain Data, we associate a corresponding domain with abstract errors

DataE = Data | Error

where by Error we denote the set of all abstract errors that are generated by our programs. Consequently every partial operation

op : $Data_1$ x … $Data_n$ → Data

whose partiality is computable[10] may be extended to a total operation

ope : $DataE_1$ x … $DataE_n$ ↦ DataE

Of course ope should coincide with op wherever op is defined, i.e. if $d_1,…,d_n$ are not errors and op.$(d_1,…,d_n)$ is defined, then ope.$(d_1,…,d_n)$ = op.$(d_1,…,d_n)$.

The operation ope is said to be *transparent for errors* or simply *transparent* if the following condition is satisfied:

if $d_k$ is the first error in the sequence $d_1,…,d_n$, then ope.$(d_1,…,d_n)$ = $d_k$

This condition means that arguments of ope are evaluated one-by-one from left to right, and the first error (if it appears) becomes the final value of the computation.

The majority of operations on data that will appear in our models will be transparent. Exceptions are boolean operations discussed in Sec.2.6

Error-handling mechanisms are frequently implemented in such a way, that errors serve only to inform the user that (and why) program evaluation has been aborted. Such a mechanism will be called *reactive*. In some applications, however, the generation of an error results in an action, e.g. of recovering the last state of a database. Such mechanisms will be called *proactive*.

---

[10] Informally speaking a partiality of a function F is computable if we can write a procedure which given an arbitrary tuple $d_1,…,d_n$ of arguments of F will check if F.$(d_1,…,d_n)$ is or is not defined. E.g. for an array expression arr[k] we can check if the index k belongs to the index range of the array arr. From the general theory of computability we know, however, that for some functions such procedures do not exist.

A reactive mechanism may be quite easily enriched to a proactive one (see Sec.6.1.8 and Sec.12.7.6.4). Since, however, the latter is technically more complicated, in our example-language **Lingua** we shall only describe a reactive model.

A well-defined error-handling mechanism allows avoiding situations where programs stop without any explanation, or even worse — when they do not stop but generate an incorrect result without any warning to the user.

## 2.6    Three-valued propositional calculus

Tertium non datur — used to say ancients masters. Computers denied this principle.

In the Aristotelean classical logic, every sentence is either true or false. The third possibility does not exist. In the world of computers, however, the third possibility is not only possible but just inevitable. In evaluating a boolean expression such as, e.g., x/y>2 an error (see Sec.2.5) will appear if x=0.

To describe the error-handling mechanism of boolean expressions, we introduce a domain of Boolean values with an error

BooleanE = {tt, ff, ee}.

In this case, ee stands for "error", but in fact, represents either an error or an infinite computation (a looping). In this section, we assume for simplicity that there is only one error. This assumption does not disturb the generality of the model as long as all errors are handled in the same way.

Now, it turns out that the transparency of boolean operators would not be an adequate choice. To see that consider a conditional instruction[11]:

**if** x ≠ 0 **and** 1/x < 10 **then** x := x+1 **else** x := x-1 **fi**

We would probably expect that for x=0 one should execute the assignment x:=x-1. If however, our conjunction would be transparent, then the expression

x ≠ 0 **and** 1/x < 10

would evaluate to 'division-by-zero' which means that the program aborts. Notice also that the transparency of **and** implies

ff **and** ee = ee

which means that when an interpreter evaluates p **and** q, then it first evaluates both p and q — as in "usual mathematics" — and only later applies **and** to them. Such a mode is called an *eager evaluation*.

An alternative to it is a *lazy evaluation* where if p = ff, then the evaluation of q is skipped, and the final value of the expression is ff. In such a case:

ff **and** ee = ff

tt **or** ee = tt

---

[11] Here I anticipate the future syntax of **Lingua** where `Courier New` is used in order to distinguish program texts form statements expressed in MetaSoft.

A three-valued propositional calculus with the above lazy evaluation was described in 1961 by John McCarthy (in [16]) who defined boolean operators as shown in Tab. 2.6-1

| **or-m** | tt | ff | ee | | **and-m** | tt | ff | ee | | **not-m** | |
|---|---|---|---|---|---|---|---|---|---|---|---|
| tt | tt | tt | tt | | tt | tt | ff | ee | | tt | ff |
| ff | tt | ff | ee | | ff | ff | ff | ff | | ff | tt |
| ee | ee | ee | ee | | ee | ee | ee | ee | | ee | ee |

**Tab. 2.6-1 Propositional operators of John McCarthy**

To see the intuition behind the evaluation of McCarthy's operators consider the expression p **or-m** q noticing that its arguments are computed from left to right[12]:

- If p = tt, then we give up the evaluation of q (lazy evaluation) and assume that the value of the expression is tt. Notice that in this case we maybe avoid an error message that could be generated by q or an infinite computation. Of course, **or-m** is not transparent for errors.

- If p = ff, then we evaluate q, and its value — possible ee — becomes the value of the expression.

- If p = ee, then this means that the evaluation of our expression aborts or loops at the evaluation of its first argument, hence the second argument is not evaluated at all. Consequently, the final value of the expression must be ee.

The rule for **and** is analogous. Notice that McCarthy's operators coincide with classical operators on classical values (grey fields in the tables). McCarthy's implication is defined classically:

p **implies-m** q = (**not-m** p) **or-m** q

As we are going to see, not all classical tautologies remain satisfied in McCarthy's calculus. Among those that remain satisfied we have:

- associativity of alternative and conjunction,

- De Morgan's laws

and among the non-satisfied are:

- **or-m** and **and-m** are not commutative, e.g., ff **and-m** ee = ff but ee **and-m** ff = ee,

- **and-m** is distributive over **or-m** only on the right-hand side, i.e.

    p **and-m** (q **or-m** s)  =  (p **and-m** q) **or-m** (p **and-m** s) however

    (q **or-m** s) **and-m** p  ≠  (q **and-m** p) **or-m** (s **and-m** p) since

    (tt **or-m** ee) **and-m** ff = ff  and  (tt **and-m** ff) **or-m** (ee **and-m** ff) = ee

- analogously **or-m** is distributive over **and-m** only on the right-hand side,

- p **or-m** (**not** p) does not need to be true but is never false,

---

[12] The suffix "-m" stands for "McCarthy" and is used to distinguish McCarthy's operators not only from classical ones but also from the operators of Kleene, which are used in SQL.

- p **and-m** (**not** p) does not need to be false but is never true.

# 3 The applicative layer of Lingua

## 3.1 The data

Data types available in **Lingua** may be split into two categories:

- *simple data* including Booleans, numbers, and words (finite strings of characters),
- *structural data* including list, many-dimensional arrays, records, trees, and their arbitrary combinations.

Structural data may „carry" simple data as well as other structural data. That means that we may build "deep" data structures, e.g., lists that carry records of arrays. Lists and tables always carry elements of the same type whereas records are not restricted in this way.

All our data (with appropriate abstract errors) and with the corresponding constructors constitute a many-sorted algebra of data. Many-sorted algebras of data, of types, of denotations, and of syntax constitute the fundament of our denotational model.

Formally the data domains in **Lingua** are defined by the following set of so called *domain equations*:

| | | |
|---|---|---|
| boo | : Boolean | = {tt, ff} |
| num | : Number | — the set of all numbers with finite decimal representations |
| ide | : Identifier | — a fixed finite subset of the domain $\text{Alphabet}^{c+}$ |
| wor | : Word | = {'}$\text{Alphabet}^{c*}${'} |
| lis | : List | = $\text{Data}^{c*}$ |
| arr | : Array | = Number $\Longrightarrow$ Data |
| rec | : Record | = Identifier $\Longrightarrow$ Data |
| dat | : Data | = Boolean \| Number \| Word \| List \| Array \| Record |

The symbols boo, num, ide etc. which precede our equations are metavariables that will run over the corresponding domains in further definitions. This is just another notational convention.

The domain Boolean consist of only two elements that represent "truth" and "false". The domains Alphabet, Number and Identifier, are the parameters of our model which means that they may differ from one implementation to another.

The Alphabet is a finite set of characters (except quotation marks), while Identifier is a finite fixed set of non-empty strings over Alphabet.

A word is a finite string (possibly empty) of the elements of Alphabet closed between apostrophes. The latter is necessary to distinguish between metavariables like boo or Boolean from metaconstants that "denote themselves" like 'division-by-zero'.

A list is a finite sequence (possibly empty) of arbitrary data.

An array is a mapping from numbers to data, and a record is a mapping from identifiers to data.

A data is a boolean, a number, a word, a list, an array or a record. Notice that identifiers are not included in data. They have been introduced only to define the domain of records. Identifiers that appear in records are called *record attributes*.

As we see, the four last equations have a recursive character, and therefore the existence of a solution of our set of equations is not evident. However, such a solution exists and is (in a sense) unique[13] which may be proved on the ground of the theory of so-called *chain-complete partially ordered sets* (Sec. 2.7 of [14]).

It is to be emphasized in this place that the domain of data, and all of its subdomains, are larger than the corresponding sets of numbers, words, lists etc. that can actually appear when executing the programs of **Lingua**:

1.  All "executable" data are restricted in their size — this is formalized be introducing a universal predicate oversized defined for all data.

2.  For any given list or array all its elements must be of the same type (see Sec.3.2).

3.  The domain of each array must be of the form {1,…,n}, i.e. must be a set of consecutive positive integers starting from 1.

The constructors of data are defined in such a way that all reachable data (cf. Sec.2.2) satisfy 1, 2 and 3.

The idea of defining the domains of data as oversets of reachable domains is a mathematical technique which makes our domain equations simpler. As a matter of fact, such a technique is well known in general mathematics. E.g. in the mathematical analysis we deal with the set of all numbers, i.e. with possibly infinite decimal representations, although in the engineering practice such numbers will never appear.

## 3.2    Composites, transfers, yokes, types and values

Every data in **Lingua** has a type. Types describe properties of data but represent entities which can be constructed and modified, independently of data. Our mechanism of types allows programmers to define their own types for future use either in defining new types or in declaring variables[14].

As we are going to see, types are pairs consisting of a *body* and a *yoke*. Every type is associated with a set of data of that type called the *clan of the type.*

Intuitively a body describes an "internal structure of a data" — e.g., indicates that a data is a number, a list or a record — and formally is a combination of tuples and mappings. The domain equation that defines the domain of bodies is the following[15]:

bod : Body    =

    {('Boolean')} | {('number')} | {('word')} |    (*simple bodies*)

    {'L'} x Body |    (*list bodies*)

    {'A'} x Body |    (*array bodies*)

    {'R'} x (Identifier $\Rightarrow$ Body)    (*record bodies*)

---

[13] It is unique in the sense that by the solution of such an equation we mean its least solution.
[14] Technical details in Sec. 5.2 of [14].
[15] This is again a recursive equation (as it was the case of data-domain equations) and again its unique solution exists.

The bodies of simple data are one-element tuples of metaconstants: ('Boolean'), ('number') or ('word'). The bodies of lists and arrays are respectively of the form ('L', bod) or ('A', bod) where the body bod is shared by all the elements of a list or of an array and where the *initials* 'L' and 'A' indicate that we are dealing with a list or with an array respectively.

A record body is of the form ('R', body-record) where body-record is a metarecord of bodies such as, e.g.:

| | |
|---|---|
| Ch-name ; | ('word'), |
| fa-name ; | ('word'), |
| birth-year ; | ('number'), |
| award-years ; | ('A', ('number')), |
| salary ; | ('number'), |
| bonus ; | ('number') |

The words on the left-hand-side of semicolons are identifiers called *attributes*. The first three attributes and the last two have simple bodies, whereas the fourth one — an array body. For the sake of further discussion, the body defined above will be referred to as employee.

With every body bod, we associate a set of data with that body called *the clan of the body* and denoted by CLAN-Bo.bod. The function CLAN-Bo is defined inductively relative to the structure of bodies. E.g., the set CLAN-Bo.employee contains records with numbers, words, and one-dimensional number arrays assigned to the respective attributes.

Next important concept from the "world" of data and types is a *composite* which is a pair (dat, bod) consisting of a data and its body such that:

dat : CLAN-Bo.bod

Composites are created in the course of the data-expressions evaluation (Sec.3.4). All data operations in **Lingua** are defined as operations on composites which permits to describe the mechanism of checking if the arguments "delivered" to an operation are of an appropriate type. E.g., if we try to put a word on a list of numbers, the corresponding operation will generate an error message.

Having defined composites, we can define *transfers* and *yokes*. Transfers are one-argument functions that transform composites or errors into composites or errors and *yokes* are transfers with Boolean composites as values. By a *Boolean composite* we mean (tt, ('Boolean')) or (ff, ('Boolean')). Transfers may also assume abstract errors as values.

Mathematically yoks are close to one-argument predicates on composites[16]. An example of a yoke that describes a property of composites whose body is employee may be the following inequality:

salary + bonus < 10000,

This yoke is satisfied whenever its (unique) argument is a record composite with (at least) the attributes salary and bonus, and the data corresponding to these attributes satisfy the corresponding inequality. In this example

---

[16] They "are closed to predicates" rather than simply "are predicates" since they assume as values composites and abstract errors rather than just Boolean values tt and ff. Consequently their logical constructors **and**, **or** and **not** are not the classical constructors but three-valued constructors of a calculus defined by John McCarthy (Sec. 2.6).

salary + bonus

is a transfer which is not a yoke. It transforms record composites into number composites.

Yokes, as defined above, appear in SQL and are called *integrity constraints*. As a matter of fact, they have been introduced them into our model to cope with SQL data (for details see Sec.12 of [14]).

Transfers have merely a technical role. We need them only to define an algebra where yokes may be created. With every transfer we associate its clan:

CLAN-Tr.tra = (com | tra.com = (tt, ('Boolean'))}

which consists of composites that satisfy that transfer. Of course, the clans of transfers which are not yokes, are empty. By TT we denote the transfer that yields (tt, ('Boolean')) for any composite.

A pair that consists of a body and a yoke is called a *type*. However, for technical reasons, types are defined as pairs consisting of a body and an arbitrary transfer. With every type typ = (bod, tra) we associate its *clan* which is the set of such composites whose data belong to the clan of the body and which satisfy the transfer. Formally:

CLAN-Ty.(bod, tra) = {(dat, bod) | dat : CLAN-Bo.bod **and** (dat, bod) : CLAN-Tr.tra}

The last concept associated with data and types is a *value,* also called *typed data.* A value is a pair (dat, typ), i.e. (dat, (bod, tra)), which we sometimes write as ((dat, bod), tra). As we see, a value may be regarded, either as a pair *data-type* or as a pair *composite-transfer.*

For technical reasons we also allow *pseudo-values* of the form (Ω, typ), where Ω is an abstract object called a *pseudo-data.*

Values are assigned in memory states to the identifiers of variables. Variable declarations assign pseudo values to variables and initializing assignments replace Ω by a data.

An assignment instruction — i.e., an instruction that assigns values to variables (see Sec.4.2) — may only change the data assigned to a variable, and in some special cases its body, but never its yoke. To change a yoke, we have to use a special yoke-oriented instruction.

Summing up, the list of domains that are associated with data and their types in **Lingua** is the following

| dat | : Data | = … (the definition in Sec.3.1) |
|-----|--------|--------------------------------|
| bod | : Body | = … (the definition above in this section) |
| com | : Composite | = {(dat, bod) | dat : CLAN-Bo.bod} |
| com | : BooComposite | = {(boo, ('Boolean')) | boo : Boolean} |
| tra | : Transfer | = (Composite | Error) ↦ (Composite | Error) |
| yok | : Yoke | = (Composite | Error) ↦ (BooComposite | Error) |
| val | : Value | = {(dat, typ) | dat = Ω **or** dat : CLAN-Ty.typ} |

In these domains:

- *data* are the data processed by programs,

- *bodies* are objects that describe "internal structures" of data,

- *composites* are pairs consisting of a data and its body; as we are going to see *data-expressions* evaluate to composites,

- *transfers* are one-argument functions on composites and errors,

- *yokes* are transfers that return Boolean composites or abstract errors,

- *types* are pairs that consist of a body and a transfer (in fact a yoke); as we are going to see *type expressions* evaluate to types, and in memory states types are assigned to *type constants*,

- *values* are pairs consisting of a data and its type; in states, values are assigned to *variable identifiers*.

Similarly, as in many programming languages (although not in all of them), types in **Lingua** have been introduced for four reasons:

1. to define a type of a variable when it is declared, and to assure that this type remains unchanged (with some exceptions)[17] during program executions,

2. to ensure that a data which is assigned to a variable by an assignment is of the type consistent with the declared type of that variable,

3. to ensure that a similar consistency takes place when sending actual parameters to a procedure or when returning reference parameters by a procedure,

4. to ensure that in evaluating an expression, an error message is generated whenever data "delivered" to that expression are of an inappropriate type, e.g., when we try to add a word to a number or to put a record to a list of arrays.

## 3.3    Expressions in general

Expressions are syntactic object and their *denotations*, i.e. their semantic meanings, are functions from states to composites (*data expressions*), to transfers (*transfer expressions*) or to types (*type expressions*). In order to define all these concepts we have to start with the definition of a *state*:

| sta | : State | = Env x Store | (state) |
|-----|---------|---------------|---------|
| env | : Env | = TypEnv x ProEnv | (environment) |
| sto | : Store | = Valuation x (Error | {'OK'}) | (store) |
| vat | : Valuation | = Identifier $\Longrightarrow$ Value | (valuation)[18] |
| tye | : TypEnv | = Identifier $\Longrightarrow$ Type | (type environment) |
| pre | : ProEnv | = Identifier $\Longrightarrow$ Procedure | Function | (procedure environ- |

ment)[19]

| env | : Environment | = TypEnv x ProEnv |
|-----|---------------|-------------------|

As we see, states are binding identifiers to values, to types, to procedures, and to functions (functional procedures) and besides may store an error in a "dedicated register". If a state does not carry an error then this register stores the word 'OK'. Every state is therefore a tuple of the form:

(env, (vat, err))    where err : Error | {'OK'}

---

[17] These exceptions take place e.g. when we add a new attribute to a record or to a database table or if we remove such attribute.

[18] The metavariable running over valuations is "vat" since "val" has been reserved for values.

[19] The domains Procedure and Function will be defined in Sec. ???

Having defined states we can define the domains of expression denotations of three categories:

| | | |
|---|---|---|
| DatExpDen | = State → Composite | Error | (data-expressions denotations) |
| TraExpDen | = State ↦ Transfer | Error | (transfer-expressions denotations) |
| TypExpDen | = State ↦ Type | Error | (type-expressions denotations) |

The denotations of data expressions are partial functions which is due to the fact that such expressions may include functional-procedure calls[20]. The remaining denotations are total functions.

The three domains are the carriers of an *algebra of expression denotations* from which a syntactic *algebra of expressions* is derived (as sketched in Sec.2.4) with the carriers DatExp, TraExp, TypExp. This leads to three functions of semantics which constitute a homomorphism between our two algebras.

| | | |
|---|---|---|
| Sde | : DatExp | ↦ DatExpDen |
| Stre | : TraExp | ↦ TraExpDen |
| Ste | : TypExp | ↦ TypExpDen |

## 3.4    Data expressions

Data expressions evaluate to composites or errors. With every operation on data, we associate two constructors: of data-expression denotations and of data expressions. In this way, we define two mutually similar algebras and then a homomorphism between them. This homomorphism is unique, is implicit[21] in the definitions of both algebras and constitutes the semantics of data expressions. In this section, I show just one example of a syntactic constructor and of the corresponding semantic clause.

Consider the data operation of the numeric division divide and its syntactic counterpart "/". The clause of our grammar (Sec.5.1.1) that corresponds to the syntactic constructor is

(DatExp / DatExp)

where the characters (,) and / belong to the syntax of the language and DatExp is a metavariable in our equational grammar.

In the sequel instead of dealing directly with grammatical clauses, I shall write them in the form of a *syntactic scheme*. In the present case, such a scheme of an expression with the division is of the form

(dae-1 / dae-2),

where dae-1 and dae-2 are metavariables denoting data expressions. I write them in Courier New to indicate that they belong to the word of syntax.

The corresponding clause of the definition of semantics is shown below. The syntactic argument is closed in square brackets.

---

[20] Functional procedures may loop indefinitely and since this is not a computable property we cannot expect to have an error message in that case.
[21] What "implicit" means in this case, will be explained in Sec.5.3.

Sde.[`(dae-1 / dae-2)`].sta =

   **let**

      (env, (val, err)) = sta

   err ≠ 'OK'              ➜ err

   Sde.[`dae-i`].sta = ?    ➜ ?             for i = 1,2

   **let**

      num-i = Sde.[`dae-i`]. (env, (val, err))    for i = 1,2

   num-i : Error         ➜ num-i       for i = 1,2

   **let**

      (dat-i, bod-i) = num-i          for i = 1,2

   bod-i ≠ ('number')    ➜ 'number-expected'   for i = 1,2

   dat-2 = 0            ➜ 'division-by-zero'

   **let**

      dat-3 = divide(dat-1, dat-2)

   oversized.dat-3        ➜ 'overflow'

   **true**               ➜ (dat-3, ('number'))

In the above definition the clause

   Sde.[`dae-i`].sta = ?  ➜ ?  for i = 1,2

stands for

   Sde.[`dae-1`].sta = ?

   Sde.[`dae-2`].sta = ?

and analogously for all similar clauses. Intuitively our definition should be read as follows:

1. If the input state carries an error, then this error becomes the final result of the computation.

2. Otherwise, we evaluate both component expressions, and if one of these evaluations does not terminate, then (of course) the whole computation does not terminate.

3. Otherwise, we check the bodies of both resulting composites and if one of them is not ('number'), then an appropriate error is generated.

4. Otherwise, we check if the second argument of the division is zero, in which case an error is generated.

5. Otherwise, we check if the result of the division is not oversized in which case an error is generated[22].

6. Otherwise, the result of division becomes part of the resulting composite.

---

[22] In our definitions this part of procedure is described in an abstract way, but the implementation does not need to preform it literally, i.e. by first dividing the given numbers and only then checkig, if that was possible. In an implementation a programmable solution should be chosen.

## 3.5    Transfer expressions

Transfer expressions evaluate to transfers or errors. Transfers are one-element functions from composites to composites. Transfers that evaluate to Boolean composites, i.e. to composites of the form (tt, ('Boolean')) or (ff, ('Boolean')) are called yokes. Since transfers are not usual in programming languages, a few examples may be in order, to better understand the nature of transfer expressions. In their descriptions, the "current composite" means the composite which is the (only) argument of the transfer.

| | |
|---|---|
| `273` | — the resulting composite is (273, ('number')) independently of the current composite, |
| `record.price` | — if the current composite carries a record with an attribute `price` of the body ('number') and data dat, then the resulting composite is (dat, ('number')), and otherwise is an error. |
| `all-list number ee` | — this is a yoke; if the current composite does not carry a list, then an error is generated, otherwise, if it is a list of numbers then the resulting composite is (tt, ('Boolean')), and otherwise, it is (ff, ('Boolean')), |
| `record.price + record.vat < 1000` | — this is a yoke; if the current composite does not carry an appropriate record, then error and otherwise, if the sum of data assigned to `price` and `vat` is less than 1000, then (tt, ('Boolean')), and otherwise (ff, ('Boolean')) |

Now let us consider a transfer expression with the asyntactic scheme

> `all-list tre ee`.

Such a transfer expressions is satisfied if all elements of a current list satisfy the transfer `tre`. The corresponding clause of the definition of semantics is the following:

Stre.[**all-list** tre **ee**].com =

    com : Error                  ➔ com

    sort.com ≠ 'L'             ➔ 'list-expected'

      **let**

        ((dat-1,…,dat-n), ('L', bod)) = com    (list elements always have the same body)

        com-i = Stre.[tre].(dat-i, bod)           for i = 1;n

      com-i : Error          ➔ com-i    for i = 1;n

      **not** com-i : BooComposite   ➔ 'a-yoke-expected'

      ($\forall$ i = 1;n) com-i = (tt, ('Boolean')) ➔ (tt, ('Boolean'))

      **true**                  ➔ (ff, ('Boolean'))

This definition may be intuitively read as follows:

1.  If the current composite is an error, then the result is this error.

2. Otherwise, if the current composite does not carry a list, then an error is signalized.

3. Otherwise, the transfer Stre.[tre] is applied to composites created from the data dat-i of the list and the "internal body" bod of the list. Notice that lists carry data, rather than composites.

4. If one of these composites is an error, then the first such an error is the result of the computation.

5. If one of these composites is not a Boolean composite, then an error is generated.

6. If all resulting composites are (tt, ('Boolean')), then the resulting composite is (tt, ('Boolean')), and otherwise, it is (ff, ('Boolean')).

In the sequel by TT, we denote a transfer which is always satisfied.

## 3.6    Type expressions

Type expressions evaluate to types or errors. E.g., the denotation of the following type expression:

```
record-type

    Ch-name        as word,

    fa-name        as word,

    birth-year     as number,

    award-years    as number-array ee

    salary         as number

    bonus          as number

ee
```

is a function on states that creates a record type or generates an error. This expression refers to two built-in types word and number and one user-defined type number-array (arrays of numbers).

Now consider an example of a syntactic scheme of an operation that creates a one-attribute record type:

```
record-type ide as tex ee
```

where ide is an identifier and tex is a type expression. The corresponding semantic clause is the following:

Ste.[ **record-type** ide **as** tex **ee** ].sta =

    **let**

        (env, (val, err)) = sta

    err ≠ 'OK'  ➜ err

    **let**

        typ = Ste.[tex]. sta

    typ : Error ➜ num-i

    **true**        ➜ (('R', [ide/typ]), TT)

This clause is read as follows:

1. If the input state carries an error, then this error becomes the result of the computation.

2. Otherwise, we compute the type defined by `tex`, and if it is an error, then this error becomes the result of the computation.

3. Otherwise, the resulting type is the record type (('R', [`ide`/typ]), TT).

To construct a many-attribute record type we use the operation of adding an attribute to a given record type with the following syntactic scheme:

**expand-record-type** `tex-1` **at** `ide` **by** `tex-2` **ee**

and to replace a current transfer of an arbitrary type defined by `tex`, by a new transfer `tre`, we use a type expression with a scheme:

**replace-transfer-in** `tex` by `tre` **ee**

# 4 The imperative layer of Lingua

Expressions of all types belong to an *applicative layer* of **Lingua**. Their denotations use states as arguments but neither create them nor change. The latter tasks are performed by *instructions, variable declaration, procedure- and function declarations* and by *type definitions*. All of them belong to an *imperative part of the language*.

## 4.1 Some auxiliary concepts

Two new metapredicates are necessary to define the semantics of the imperative layer of our language. The metapredicate

is-error : State $\mapsto$ {tt, ff}

returns tt whenever a state carries an error. We say that body **bod-1** *is coherent* with **bod-2**, in symbols

bod-1 <u>coherent</u> bod-2

whenever:

1. bod-1 = bod-2 or

2. both bodies are record-bodies, and the set of attributes of one of them is a subset of the set of attributes of the other.

In other words, two bodies are coherent if they are identical, or if they are record bodies and one of them results from the other by adding or by removing an attribute. We also introduce an operator of inserting an error into a state:

◄ : State $\mapsto$ State

(env, (vat, err)) ◄ error = (env, (vat, error))

## 4.2 Instructions

Instructions change states, and therefore instruction denotations are partial functions from states to states:

InsDen = State → State

The partiality of these denotations results from the fact that the execution of an instruction may be infinite. The semantics of instructions is, therefore, a function

Sin : Instruction ↦ InsDen

Contrary to expression denotations which may generate an error, instruction denotations write an error into the error register of the state. The denotations of the majority of instructions are *transparent* relative to error-carrying states, i.e., they do not change such states but only pass them to the subsequent parts of the program. However, an error may also cause an error-handling action.

The basic instruction is, of course, the *assignment* of a value to a variable identifier. The syntactic scheme of an assignment instructions is:

```
ide := dae
```

and the corresponding semantic clause is the following:

Sin.[`ide := dae`].sta =

    is-error.sta                     ➔ sta

    **let**

        ((tye, pre), (vat, 'OK')) = sta

    vat.ide = ?                 ➔ sta ◄ 'identifier-not-declared'

    Sde.[`dae`].sta = ?       ➔ ?               (an infinite execution)

    Sde.[`dae`].sta : Error    ➔ sta ◄ Sde.[`dae`].sta

    **let**

        ((dat-f, bod-f), tra) = vat.ide          (f – former)

        (dat-n, bod-n)     = Sde.[`dae`].sta        (n – new)

        com               = tra.(dat-n, bod-n)

    com : Error               ➔ sta ◄ com

    **not** bod-n coherent bod-f    ➔ sta ◄ 'no-coherence'

    **not** com : BooComposite     ➔ sta ◄ 'a-yoke-expected'

    com = (ff, ('Boolean'))      ➔ sta ◄ 'yoke-not-satisfied'

    **let**

        val-n = ((dat-n, bod-n), tra)

    **true**                    ➔ ((tye, pre), (vat[ide/val-n], 'OK'))

The denotation of an assignment changes an input state into an output state in nine steps:

1. If an input state carries an error, then this state becomes the output state.

2. Otherwise, if the identifier `ide` has not been declared, i.e., if no value or a pseudo value has been assigned to it in the valuation val, then an error message is loaded to the error register.

3. Otherwise, if an attempt to evaluate the data expression leads to an infinite execution, then (of course) the executions of the instruction is infinite as well.

4. Otherwise, if the expression evaluates to an error, then this error is loaded to the error register of the state.

5. Otherwise, it the transit applied to the new composite returns an error, then this error is loaded to the error register.

6. Otherwise, if the composite computed from the expression has a body non-coherent with the body of the identifier's type, then an error is loaded to the error register.

7. Otherwise, if the composite computed by the transit is not Boolean, i.e. if the transit was not a yoke, then an error is loaded to the error register.

8. Otherwise, if the yoke is not satisfied, then an error message is loaded to the error register.

9. Otherwise, the new value is the new composite and the current (i.e. not changed) yoke, and this new value is assigned to the identifier `ide`.

Notice that as a consequence of the claim 6. together with the definition of the coherence of bodies (Sec.4.1) an assignment may change the body of a value assigned to a variable only if this body is a record, and only by adding or by removing an attribute to/from that record.

The remaining instructions belong to one of the following seven categories where the first four are *atomic instructions,* and the other three are *structural instructions*, i.e., instructions composed of other instructions and expressions:

1. the replacement of a yoke assigned to a variable by another one
   **yoke** `ide := tre`,

2. the empty instruction
   **skip**,

3. the call of an imperative procedure
   **call** `ide (`**ref** `apar-r` **val** `apar-v)`
   where `apar-r` and `apar-v` are, (maybe empty) lists of identifiers called respectively *actual reference-parameters* and *actual value-parameters*,

4. the activation of an error-handling
   **if** `dae` **then** `ins` **fi**,

5. the conditional composition of instructions
   **if** `dae` **then** `ins-1` **else** `ins-2` **fi**,

6. the loop
   **while** `dae` **do** `ins` **od** ,

7. the sequence of instructions
   `ins-1 ; ins-2`.

In the yoke-replacement instruction, the new value of the identifier `ide` gets the old composite but a new transfer. This transfer must be satisfied by the current composite[23].

The empty instruction **skip** is needed to make functional-procedure declarations sufficiently universal; this will be seen in Sec.4.4.

---

[23] This instruction has been introduced mainly for the sake of SQL tables discussed in [14].

The discussion of procedures is postponed to Sec.4.4

The error handling is activated if the current state carries an error, i.e. a word, that is equal to the word that the data-expression `dae` evaluates to. If this happens, the "internal" instruction `ins` is executed for a state that results from the initial state where the error has been replaced by 'OK'[24].

The semantics of the three remaining categories of instruction is usual with the exception that in the last two cases an expression may generate an error message. In such a case that error is stored in the error register of the state.

## 4.3    Variable declaration and type definitions

*Variable-declaration denotations* are total functions that map states into states:

VarDecDen = State $\longmapsto$ State

assigning types to identifiers and leaving their data undefined. More formally, they assign pseudo-values (Sec.3.2)*,* i.e. pairs of the form $(\Omega, \text{typ})$. The syntactic scheme of a single declaration is of the form:

**let** ide **be** tex **tel**

Variable declarations are similar to assignments with the difference that for a declaration an error 'identifier-not-free' is signalized whenever the identifier `ide` is bound in the input state. It means that a variable may be declared in a program only once. During future program-execution its value may be changed only by changing:

- the composite of the value by an assignment instruction,
- the yoke of the value by a yoke-replacement.

Type definitions are of the form

**set** ide **as** tex **tes**

and their denotations are similar to those of variable declarations, i.e.

TypDefDen = State $\longmapsto$ State

with the difference that instead of assigning a pseudovalue to a variable identifier in a valuation they assign a type to a type-constant identifier in a type environment.

An identifier that is bound to a type in a state is called a *type constant*. Notice that "a constant" rather than "a variable" since a type once assigned to an identifier, cannot be changed in the future (an engineering decision).

Similarly as in the case of assignments, also type definitions, and variable declarations may be combined sequentially using a semicolon constructor.

## 4.4    Procedures

Procedures in **Lingua** may be *imperative* or *functional*. The former are functions that take two lists of actual parameters — value parameters and reference parameters — and return partial

---

[24] For details see Sec.6.1.8 of [14].

functions on stores[25]. Functional procedures take only value parameters and return partial functions from states to composites or errors:

ipr : ImpPro = ActPar x ActPar ↦ Store → Store

fpr : FunPro = ActPar ↦ State → (Composite | Error)

In these equations, ActPar is a domain of *actual-parameter lists* defined by the domain equation:

apa : ActPar = () | Identifier | ActPar x ActPar

As we see, actual-parameter lists are finite (maybe empty) sequences of identifiers. In turn, formal-parameter lists that appear in procedure declarations are finite (maybe empty) sequences of pairs consisting of an identifier and a type-expression denotations:

fpa : ForPar = () | Identifier x TypExpDen | ForPar x ForPar

Returning to procedures notice that we do not talk here about procedure denotations but about procedures as such since they are "purely denotational" concepts. In other words, they do not have syntactic counterparts. At the level of syntax, we have only *procedure declarations* and *procedure calls* which, of course, have their denotations.

A syntactic scheme of an imperative-procedure declaration is of the following form (the carriage returns are of course syntactically irrelevant):

```
proc ide (ref fpar-r val fpar-v)

    pro

end proc
```

where `pro` is a program (see later) and `fpar-r` and `fpar-v` are the lists of respectively formal reference-parameters and formal value-parameters. A syntactic example of a list of formal parameters may be as follows:

```
(val age, weight as number, name as word

 ref patient as patient-record)
```

Expressions different from single-identifier-expressions are not allowed as value parameters since such a solution would complicate the model as well as program-construction rules (an engineering decision).

If we want to declare a group of mutually recursive procedures then we use a *multiprocedure declaration* of the form:

```
begin multiproc

    ipd-1;

    ipd-2;

    …

    ipd-n

end multiproc
```

---

[25] The fact that procedures transform stores rather than states is a technique that allows to avoid selfapplication of procedures, i.e. a situations where a procedure takes itself as an actual parameter. Of course, procedure calls are instructions and therefore they transform states into states.

where the `ipd`'s are imperative-procedure declarations. Intuitively this means that these procedure declarations have to be elaborated "as a whole", rather than one after another (details in Sec.7.4 of [14]).

The syntactic scheme of a functional-procedure declaration is of the form :

**fun** `ide (fpar)`

   `pro`

**return** `dae` **as** `tex`

A call of a functional procedure declared in this way first executes the program `pro` and then evaluates the data expression `dae` in the output state of the program. If the composite generated by that expression is of the type defined by the type expression `tex`, then this composite becomes the result of the call of the function. Otherwise, an error is signalized.

In particular, the program in a functional-procedure declaration may be the trivial instruction **skip** — which "does nothing" — and the exporting expression may be a single identifier.

The syntactic schemes of an imperative-procedure call and a functional-procedure call are respectively:

**call** `ide (`**ref** `apar-r` **val** `apar-v)`    — imperative-procedure call

`ide (apar-v)`                — functional-procedure call

Notice that the second call has no reference parameters since functional procedures do not have any *side-effects* — they do not modify a state (an engineering decision).

Procedures discussed above accept as parameters only variable identifiers, i.e., identifiers that bind values. All types and procedures defined in the "main" program <u>before</u> (see Sec.4.4) the declaration of a procedure are visible in the body of this procedure, and therefore they do not need to be passed as parameters (an engineering decision).

In the version of **Lingua** described in the present paper procedures cannot take other procedures as parameters. However, it is shown in [14] (Sec. 7.6) how to overcome this restriction by constructing a hierarchy of procedures that can take as parameters only procedures of a lower rank than themselves. This construction protects procedures from taking themselves as parameters which would lead to non-denotational models (a mathematical decision).

## 4.5    The execution of a procedure call

In the descriptions of procedure mechanisms, we shall use some concepts having to do with the fact that procedures are created when they are declared and are executed when they are called. In respect to that, we shall talk about states (and their components) of a *declaration-time* and of a *call-time* respectively[26]. Traditionally by a *procedure body,* we shall mean the program that is executed when a procedure is called.

As has been already announced, in **Lingua-2** there will be no global variables in procedures (an engineering decision)[27]. The intention is that the head of a procedure-call describes explicitly and completely the communication mechanisms between a procedure and the hosting program. That solution may seem restrictive but — in my opinion — guarantees a better understanding of program functionality by programmers and definitely simplifies program-construction rulers.

---

[26] These ideas, similarly to a few others, have been borrowed from M. Gordon [15]

[27] If we would like to introduced global variables, we should define the local store of a procedure call as a modification of its global store.

An execution of a procedure call may be symbolically split into four stages illustrated in Fig. 4.5-1. (technical details in Sec.7.3 of [14]).

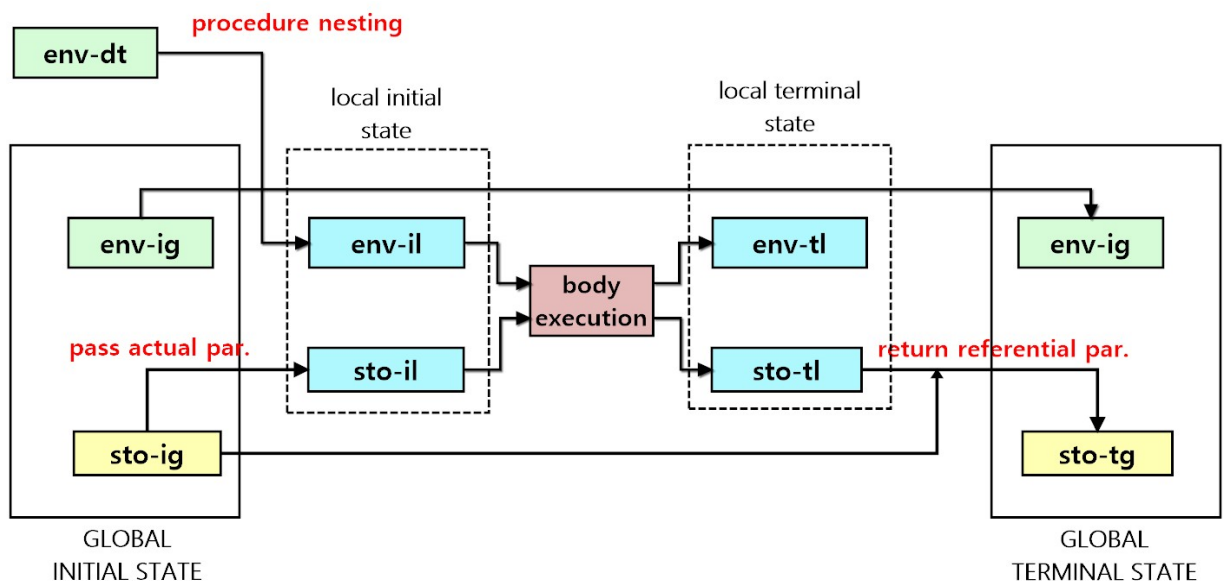

**Fig. 4.5-1 The execution of a procedure call**

1. **The inspection of an *initial global state*** — that state consists of:

   a.   an *initial global environment* env-ig,

   b.   an *initial global store* sto-ig = (vat-ig, err)

   If err ≠ 'OK', then the initial global state is returned by procedure call and therefore becomes the terminal global state. In the opposite case, an initial local state is created.

2. **The creation of an *initial local state*** — that state consists of:

   a.   *initial local environment* env-il created from the <u>declaration-time environment</u> by nesting in it the called procedure; this nesting is necessary to enable recursive calls,

   b.   *initial local valuation* vat-il covering only formal parameters with assigned values of corresponding actual parameters; to get the latter values, we refer to initial global valuation val-ig.

3. **The transformation of the local initial state** by executing the procedure body. If this execution terminates, then the local terminal state consists of:

   a.   *terminal local environment* env-tl,

   b.   *terminal local store* sto-tl = (val-tl, err-tl).

   If err-tl ≠ 'OK', then a global terminal state is created from the initial global-state by loading to it err-tl. Notice that in this case, the terminal local-environment and terminal local store are "abandoned". Otherwise, the terminal global state is created.

4. **The  creation of the *terminal global state*** — that state consists of:

    a. *initial global environment* env-ig; notice that terminal local environment env-tl is "abandoned",

    b. *terminal global store* sto-tg created from initial global store sto-ig by "returning" to it the values of formal referential parameters (stored in sto-tl) and assigning them to the corresponding actual referential parameters.

Notice that initial local environment "inherits" all types and procedures from the declaration-time environment. Procedure body may keep in it its own local environment types and procedures, but after the completion of the call, they cease to exist, since the hosting program returns to the initial global environment.

It is to be underlined that the procedure body may access only that part of the environment which was created <u>before the procedure declaration</u>.

Of a similar character is the local valuation that is created only in procedure execution-time, although in this case the values or reference-parameters stored in it are eventually returned to the terminal global valuation.

Summarizing visibility rules concerning procedure call:

1. the only variables visible in procedure-body are formal parameters plus variables local to the body (declared in it),

2. the only types and procedures visible in procedure-body are declaration-time types and procedures plus locally declared ones,

3. variables, types and procedures declared in the procedure-body are not visible outside of procedure call.

All these choices are not mathematical necessities but pragmatic engineering decisions dictated by the intention of making our model relatively simple which should contribute to the simplicity of program-construction rules and a better understanding of program-behaviour by language-users.

At the end one methodological remark. From an implementational view-point, the described mechanism of recursion requires that the initial global state is kept unchanged (memorized) during procedure-execution to recall it at the end. Consequently, the fact that a procedure may have many recursive calls means that each call should "memorize" its initial states. That mechanism is usually implemented by a stack of states. This is an iterative implementation of recursion. In our case, however, we do not need to use that method since the recursion in **Lingua** may be defined in using fixed-point recursion of **MetaSoft** (see Sec.7.3.2 in [14]).

## 4.6     Preambles and programs

Each program in **Lingua** consists of a preamble followed by an instruction. The syntactic scheme of a program is therefore of the form:

    **begin-program** pam ; ins **end-program**

where pam is a preamble.

Preambles are sequential compositions of type-constant definitions, data-variable declarations and procedure declarations. Their syntax is defined by the following grammatical clause:

pam : Preamble =

    ImpProDec        |

MultiProDec      |

FunProDec       |

TypDef          |

VarDec          |

**skip**        |

Preamble ; Preamble

Similarly to instructions also preambles contains **skip** which represent an identity state-to-state transformation. The semantics of programs and preambles are the following functions:

Spr  : Program    ↦ ProDen

Spre : Preamble   ↦ PreDen

which we define by structural induction:

Spr.[`pam ; ins`] = Spre.[`pam`] ● Sin.[`ins`]

and

| | |
|---|---|
| Spre.[`ipd`] | = Sipd.[`ipd`] |
| Spre.[`mpd`] | = Smpd.[`mpd`] |
| Spre.[`fpd`] | = Sfpd.[`fpd`] |
| Spre.[`tde`] | = Std.[`tde`] |
| Spre.[`vde`] | = Svd.[`vde`] |
| Spre.[**skip**] | = Sin.[**skip**] |
| Spre.[`pam-1 ; pam-2`] | = Spre.[`pam-1`] ● Spre.[`pam-2`] |

Intuitively the clauses for preambles are read as follows:

- the semantics of preambles applied to imperative-procedure declarations coincide with the semantics of such declarations,

- the semantics of preambles applied to multi-procedure declarations coincide with the semantics of such declarations,

- …

- the denotation of a sequential composition of preambles is a sequential composition of its denotations.

Programs with the trivial preamble **skip** — if executed "without a context" — will always generate an error (unless they are the **skip** themselves). Such programs are allowed because they may appear in procedure declarations as the bodies of procedures without locally declared objects. In turn, programs with trivial preambles and instructions at the same time are allowed in the declarations of functional procedures[28].

---

[28] Both these solutions, although in a slightly different form, have been suggested to me by Andrzej Tarlecki.

# 5  Formal definitions

## 5.1    Concrete syntax

To describe the syntax of **Lingua** we use *equational grammars* introduced in Sec.2.3. Each such grammar describes a tuple of formal languages. In our case this is the following 15-tuple:

| | |
|---|---|
| DatExp | — data expressions |
| TraExp | — type expressions |
| TypExp | — transfer expressions |
| VarDec | — variable declarations |
| TypDef | — type definitions |
| Instruction | — instructions |
| ActParameters | — the lists of actual parameters |
| ForParameters | — the lists of formal parameters |
| ProComponents | — procedure components |
| ImpProcDec | — imperative-procedure declarations |
| MprComponents | — multiprocedure components |
| MultiProcDec | — multiprocedure components |
| FunProDec | — functional-procedure declarations |
| Preamble | — preambles |
| Program | — programs |

The elements of this tuple are also called the *sorts* of a language. Each equational grammar defines uniquely a many-sorted reachable algebra where the sorts of the language are the carriers of the algebra, and where the *clauses* of the grammar (i.e. the rows in its definition, see below) correspond to the constructors of this grammar. That was roughly explained in Sec.2.3.

The equational grammar of **Lingua,** which is seen below, has been split into sections which correspond to the successive sorts of the language. It is to be remembered, however, that despite its partition, the grammar represents one set of mutually recursive equations.

The terminals of the grammar are written in `Courier New` whereas the nonterminals — in Arial.

### 5.1.1    The syntax of data expressions

Below the sets NumberS and WordS denote the sets of numbers respectively words which are *syntactically representable*, e.g., with a limited number of characters in them. The clauses with `num` and `wor` mean that each representable number or word is in the sort of data expressions. In other words, it does not need to be constructed but may be manually written as such from the keyboard. Yet another explanation may be that in a parsing tree of a program they correspond to leaves.

dae : DatExp =

**constants**

```
true | false
```
       |

```
num
```
       | (for every **num** : NumberS)

```
wor
```
       |   (for every **wor** : WordS)

**variables**

    Identifier        |

**Boolean expressions**

    (DatExp **and** DatExp)        |

    (DatExp **or** DatExp)        |

    (**not** DatExp)        |

    (DatExp < DatExp)        |

**numeric expressions**

    (DatExp + DatExp)        |

    (DatExp / DatExp)        |

**word expressions**

    DatExp **glue** DatExp        | (ambiguity)

**list expressions**

    **list** DatExp **ee**        |

    **push** DatExp **on** DatExp **ee**        |

    **top** (DatExp)        |

    **pop** (DatExp)        |

**array expressions**

    **array** DatExp **ee**        |

    **add-to-arr** DatExp **new** DatExp **ee**        |

    **change-arr** DatExp **at** DatExp **by** DatExp        |

    **arr** DatExp **at** DatExp **ee**        |

**record expressions**

    **record** Identifier **of-value** DatExp **ee**        |

    **add-atr** Identifier **of-value** DatExp **to** DatExp **ee**        |

    **rec** DatExp **at** Identifier **ee**        |

    **remove-atr** Identifier **from** DatExp **ee**        |

    **change-rec** DatExp **at** Identifier **by** DatExp **ee**        |

**conditional expression**

    **if** DatExp **then** DatExp **else** DatExp **fi**

**a functional-procedure call**

    Identifier (ActParameters)

Notice that the clause DatExp **glue** DatExp introduces ambiguity to our grammar, since, e.g. the expression `abc` **glue** `prs` **glue** `xyz` may be parsed in two different ways:

    `abc` **glue** `[prs` **glue** `xyz]` and

    `[abc` **glue** `prs]` **glue** `xyz`

This ambiguity is not harmful, however, since due to the associativity of the concatenation of words, both these parsing trees will be mapped to the same word abcprsxyz (more on that issue in Sec.2.13 of [14]).

## 5.1.2    The syntax of transfer expressions

    tre : TraExp =

**processing expressions**

| `num` | \| for every num : NumberS |
|---|---|
| `wor` | \| for every wor : WordS |
| (TraExp + TraExp) | \| |
| (TraExp / TraExp) | \| |
| **sum** (TraExp) | \| |
| **max** (TraExp) | \| |
| TraExp **glue** TraExp | \| |

**transfer-yoke expressions**

| `true` \| `false` | \| |
|---|---|
| (TraExp = TraExp) | \| |
| (TraExp < TraExp) | \| |
| **small-number** (TraExp) | \| |
| **increasing** (TraExp) | \| |
| (TraExp **and** TraExp) | \| |
| (TraExp **or** TraExp) | \| |
| (**not** TraExp) | \| |

**quantifier expressions**

| **all-list** TraExp **ee** | \| |
|---|---|
| **all-array** TraExp **ee** | \| |

**selection expressions**

| **top** | \| |
|---|---|
| **array**[TraExp] | \| |
| **record**.Identifier | \| |

**passing expression**

`value`                                      |

## 5.1.3     The syntax of type expressions

tex :TypExp =

`boolean`                                            |

`number`                                             |

`word`                                               |

Identifier                                           |

**`list-type`**  TypExp **`ee`**                          |

**`array-type`**  TypExp **`ee`**                         |

**`record-type`**  Identifier **`as`** TypExp **`ee`**          |

**`expand-record-type`**  TypExp **`at`** Identifier **`by`**  TypExp  **`ee`**  |

**`replace-transfer-in`** TypExp **`by`** TraExp **`ee`**

## 5.1.4     The syntax of variable declarations

vde : VarDec =

**`let`**  Identifier **`be`**  TypExp **`tel`**  |

VarDec ; VarDec

## 5.1.5     The syntax of type definitions

tde : TypDef =

**`set`** Identifier **`as`** TypExp **`tes`**   |

TypDef ; TypDef

## 5.1.6     The syntax of actual and formal parameters

apar : ActParameters =

**`empty-ap`**   |

Identifier    |

ActParameters , ActParameters

fpar : ForParameters =

**`empty-fp`**          |

Identifier **`as`** TypExp |

ForParameters , ForParameters

### 5.1.7      The syntax of imperative procedure declarations

ipd : ImpProcDec =

   **proc** Identifier (**val** ForParameters **ref** ForParameters)

     Program

   **end proc**

### 5.1.8      The syntax of imperative multiprocedure declarations

mpd : MultiProcDec =

   **begin multiproc**

     [ ImpProcDec ]$^{c+}$

   **end multiproc**

In this equation [ ImpProcDec ]$^{c+}$ means that ImpProcDec may be repeated an arbitrary positive number of times.

### 5.1.9      The syntax of functional procedure declarations

fpd : FunProDec =

  **fun** Identifier (ForParameters) DatExp **endfun** |

  **fun** Identifier (ForParameters)

    Program

    **return** Identifier **as** TypExp

  **and fun**

### 5.1.10     The syntax of instructions

ins : Instruction =

   Identifier := DatExp                                                |

   **yoke** Identifier := TraExp                                       |

   **skip**                                                            |

   **call** Identifier (**ref** ActParameters **val** ActParameters) |

   **if** DatExp **then** Instruction **else** Instruction **fi**      |

   **if-error** DatExp **then** Instruction **fi**                    |

   **while** DatExp **do** Instruction **od**                         |

   Instruction ; Instruction

### 5.1.11     The syntax of preambles

pam : Preamble =

   ImpProDec        |

MultiProDec    |

FunProDec      |

TypDef         |

VarDec         |

**skip**           |

Preamble ; Preamble

## 5.1.12    The syntax of programs

prg : Program =

    **begin-program** Instruction **end-program** |

    **begin-program** Preamble ; Instruction **end-program**

# 5.2    Colloquial syntax

The definition of a colloquial syntax is a very important step in the process of a language design since it makes the language more user-friendly. We free ourselves from the algebraic rigor of concrete syntax without losing anything of mathematical precision.

For **Lingua** we shall assume that colloquial syntax includes all concrete syntax which means that the use of colloquialisms is optional.

## 5.2.1    Universal rules on expressions

The following rules concern all sorts of expressions:

1. we allow spaces and carriage returns which will be removed by the restoring transformation,

2. none of the keywords `true, false,` **`if, then,`**… cannot be used as an identifier; in this case, restoring transformation does not modify a program but only generates an error message; in traditional parsers, this analysis is performed at the lexical level.

## 5.2.2    Numeric and Boolean data expressions

For both categories of expressions, we allow the omission of the "unnecessary" parentheses, which means that instead of writing `(x+(y*z))` we write `x+y*z`. However, in defining the restoring function which adds parentheses, we have to take into account that the addition and the multiplication are not associative which is due to the effect of overloading. E.g., if the maximal size of a number is 10, then

    ((-4 + 9) + 2) = 7

    (-4 + (9 + 2)) = 'overload'

The usual practice is therefore that parentheses-free expressions are evaluated from left to right in using the priorities between operations. This means that, e.g., the expression:

    `x + y + z + x*y`

is restored to

```
((x + y) + z) + (x*z)
```

## 5.2.3    Array data expression

In this category, we have four colloquialisms. The first of them concerns the constructor of an array. For instance, the colloquial expression

**array** [x, x+y,  3*y]

unfolds to the concrete expression:

**add-to-arr**                                                    (add value 3*y to the array)

   **add-to-arr**                                                 (add value x+y to the array)

      **array** x **ee**                            (create one-element array with value x)

   **new** x+y **ee**

**new** 3*y **ee**

Of course, each simple numerical expression may be replaced here by an arbitrary expression. If measurement-data is an array variable, then the colloquial expression

measurement-data.[x+1]

unfolds to concrete expression

**arr** measurement-data **at** x+1 **ee**

and

measurement-data.[x+1].[y-1]

unfolds to:

**arr arr** measurement-data **at** x+1 **ee at** y-1 **ee**

The case of adding a new element to an array may be treated analogously:

**add-to-arr** measurement-data **new** [x, x + y,  3*y] **ee**

and in the case of array modification (here we introduce a new symbol „<="):

**change-arr** measurement-data **by**

   s    <= x,

   s+1 <= x+y,

   3*p <= z-1

**ee**

which unfolds to:

**change-arr**

   **change-arr**

      **change-arr** measurement-data **at** s **by** x **ee**

      **at** s+1 **by** x+y **ee**

   **at** 3*p **by** z-1 **ee**

## 5.2.4     Record-data expression

Examples for records may be similar to these for arrays. For instance, we may assume that a colloquial expression:

```
record
    ch-name     <= 'John',
    fa-name     <= 'Smith',
    birth-date  <= 1968,
    award-years <= award-years-Smith
ee
```

corresponds to the concrete:

```
add-atr award-years      of-value award-years-Smith to
  add-atr birth-date      of-value 1968 to
    add-atr fa-name       of-value 'Smith' to
      set-record ch-name of-value 'John'
        ee
      ee
    ee
  ee
```

and a colloquial expression

```
employee.(fa-name)
```

corresponds to the concrete:

```
rec employee at fa-name ee
```

Notice that despite a similarity between selection expression from an array and from a record, there is no ambiguity since array indices are closed in bracket parenthesis and record indices in ordinary parenthesis. Therefore, if `employee` is an array variable, then the corresponding selection expression would have the form

```
employee.[fa-name]
```

## 5.2.5     Array transfer expressions

We use school rules for dropping parentheses with corresponding priorities between operations. For instance in the place of:

```
(2+value)< 10
```

we write

```
2+value < 10
```

In the place of

```
get-from-array x+1 ee
```

we write

**`array.`**`[x+1]`

It is to be recalled that in this case **`array`** is not an array variable — as, e.g. in the expression `measurement-data.[x+1]` — but a keyword that means that the input composite of this transfer should carry an array and our expression selects from this array an element with index `x+1`.

## 5.2.6    Record type expressions

In this case, we introduce colloquialisms analogous as in data expressions. For instance:

**`record-type`**

    `ch-name`     **`as`** `string,`

    `fa-name`     **`as`** `string,`

    `birth-date` **`as`** `number,`

    `award-years` **`as`** **`array-of`** `number` **`ee`**

**`ee`**

unfolds to a concrete written employing **`record-of`**  and **`expand-record`**.

## 5.2.7    Record transfer expression

in this case similarly as for arrays we write:

**`record.`**`fa-name`

instead of

**`get-from-record`** `fa-name` **`ee`**

In the first expression, **`record`** is a keyword similarly to **`array`** in case of arrays.

## 5.2.8    Type expressions

In the majority of programming languages yokes do not appear in the definitions of types, hence in such cases, the concrete syntax of type definitions would be of the form:

**`set-type`** `TypExp` **`with`** `true` **`ee`**

for example:

**`set-type`** **`array-of`** `number` **`ee`** **`with`** `true` **`ee`**

In that case, the corresponding colloquial expression would be

**`set-type`**

    **`array-of`** `number` **`ee`**

**`ee`**

The general rule is such that if the yoke is the constant `true,` then we drop the whole phrase „**`with`** `true`".

In the case of record types, we introduce colloquialisms that allow describing yokes and bodies in one expression. For instance, we write:

**`record-type`**

```
    ch-name       as string,
    fa-name       as string,
    birth-date    as number with small-number,
    award-years as array-of number with small-number ee
  ee
```

which means

```
  type
  record-of
    ch-name       as string,
    fa-name       as string,
    birth-date    as number,
    award-years as array-of number ee
    ee
  with
  small-number(record.birth-date) and
  all-of-array record.award-years with small-number(value) ee
  ee
```

The yoke of this record type is satisfied if the following three conditions are satisfied:

1. input composite carries a record with at least two attributes `birth-date` and `award-years`.

2. a small number is assigned to the attribute `birth-date`,

3. an array of small numbers is assigned to `award-years`.

The remaining information about the record is included in the type expression.

### 5.2.9     Procedure calls

We allow grouping parameters into lists of variables associated with a common type as in the following example:

```
  proc name(
    val w,z as real
    ref x,y as real a,b,c as employee
          )
```

## 5.3     Semantics — general remarks

As was already mentioned in the Introduction, the semantic clauses of **Lingua** are not going to be listed in this paper. Although they should be present in any "real" manual, I skip them to keep the paper of an acceptable size. Readers who are interested in technical details should refer to [14]. In that case, however, they will not find the definition of semantics in an explicit form, e.g., as in Sec. 4.2. Since [14] was aimed at showing how to develop a language in using denotational techniques, the definition of semantics is implicit in the correspondence between the constructors of two basic algebras: the algebra of syntax and the algebra of denotations. To explain that consider two such constructors which correspond to the assignment instruction. The syntactic constructor takes an identifier and a data expression and creates an instruction:

syn-assign : Identifier x DatExp ⟼ Instruction

syn-assign.(`ide, dae`) = `ide := dae`

whereas the corresponding denotational constructor takes an identifier and a data-expression denotation and creates a denotation of an instruction:

assign : Identifier x DatExpDen ⟼ InsDen

assign.(ide, ded).sta =

    is-error.sta                          ➔ sta

    **let**

        ((tye, pre), (vat, 'OK')) = sta

    vat.ide = ?                      ➔ sta ◄ 'identifier-not-declared'

    ded.sta = ?                      ➔ ?         (an infinite execution)

    ded.sta : Error                ➔ sta ◄ ded.sta

    **let**

        ((dat-f, bod-f), tra) = vat.`ide`              (f − former)

        (dat-n, bod-n)     = ded.sta              (n − new)

        com              = tra.(dat-n, bod-n)

    com : Error                  ➔ sta ◄ com

    **not** bod-n coherent bod-f     ➔ sta ◄ 'no-coherence'

    **not** com : BooComposite     ➔ sta ◄ 'a-yoke-expected'

    com ≠ (tt, ('Boolean')        ➔ sta ◄ 'yoke-not-satisfied'

     **let**

        val-n = ((dat-n, bod-n), tra)

    **true**                         ➔ ((tye, pre), (vat[ide/val-n], 'OK'))

From these two definitions we now synthesize the semantic clause to be seen in Sec. 4.2.

# Index